\def\year{2020}\relax
\documentclass[letterpaper]{article} % DO NOT CHANGE THIS
\usepackage{aaai20}  % DO NOT CHANGE THIS
\usepackage{times}  % DO NOT CHANGE THIS
\usepackage{helvet} % DO NOT CHANGE THIS
\usepackage{courier}  % DO NOT CHANGE THIS
\usepackage[hyphens]{url}  % DO NOT CHANGE THIS
\usepackage{graphicx} % DO NOT CHANGE THIS
\usepackage{graphicx}  % DO NOT CHANGE THIS
\usepackage[utf8]{inputenc}

\usepackage{colortbl,array} % für farbige cells
\usepackage{multirow,bigdelim}

\usepackage{amsmath}
\usepackage{placeins}

\usepackage{stfloats}

\usepackage{amsfonts}

\usepackage{booktabs}
\usepackage{multirow}

\usepackage[utf8]{inputenc}
\usepackage{graphicx}

\usepackage{tabularx}
\newcolumntype{L}{>{\raggedright\arraybackslash}X}

\title{High Tissue Contrast MRI Synthesis Using Multi-Stage Attention-GAN for Glioma Segmentation}

\author{Mohammad Hamghalam,\textsuperscript{\rm 1,2}
	Baiying Lei,\textsuperscript{\rm 1}\thanks{Corresponding author. This work was supported partly by National Natural Science Foundation of China (Nos. 61871274, 61801305, and 81571758), National Natural Science Foundation of Guangdong Province (No. 2017A030313377), Guangdong Pearl River Talents Plan (2016ZT06S220), Shenzhen Peacock Plan (Nos. KQTD2016053112051497 and KQTD2015033016104926), and Shenzhen Key Basic Research Project (Nos. JCYJ20170413152804728, JCYJ20180507184647636, JCYJ20170818142347251, and JCYJ20170818094109846).}
	Tianfu Wang\textsuperscript{\rm 1}\\ % All authors must be in the same font size and format. Use \Large and \textbf to achieve this result when breaking a line
	\textsuperscript{\rm 1}National-Regional Key Technology Engineering Laboratory for Medical Ultrasound, \\Guangdong Key Laboratory for Biomedical Measurements and Ultrasound Imaging,\\ School of Biomedical Engineering, Health Science Center, Shenzhen University,\\ Shenzhen, China, 518060,\\
	\textsuperscript{\rm 2}Faculty of Electrical, Biomedical and Mechatronics Engineering, Qazvin Branch,\\ Islamic Azad
	University, Qazvin, Iran\\
	m.hamghalam@gmail.com, leiby@szu.edu.cn, tfwang@szu.edu.cn
}

\begin{document}

\maketitle

\begin{abstract}

Magnetic resonance imaging (MRI) provides varying tissue contrast images of internal organs based on a strong magnetic field. Despite the non-invasive advantage of MRI in frequent imaging, the low contrast MR images in the target area make tissue segmentation a challenging problem.
This paper demonstrates the potential benefits of image-to-image translation techniques to generate synthetic high tissue contrast (HTC) images. Notably, we adopt a new cycle generative adversarial network (CycleGAN) with an attention mechanism to increase the contrast within underlying tissues. The attention block, as well as training on HTC images, guides our model to converge on certain tissues.
To increase the resolution of HTC images, we employ multi-stage architecture to focus on one particular tissue as a foreground and filter out the irrelevant background in each stage. This multi-stage structure also alleviates the common artifacts of the synthetic images by decreasing the gap between source and target domains. 
We show the application of our method for synthesizing HTC images on brain MR scans, including glioma tumor. We also employ HTC MR images in both the end-to-end and two-stage segmentation structure to confirm the effectiveness of these images. The experiments over three competitive segmentation baselines on BraTS 2018 dataset indicate that incorporating the synthetic HTC images in the multi-modal segmentation framework improves the average Dice scores 0.8$\%$, 0.6$\%$, and 0.5$\%$ on the whole tumor, tumor core, and enhancing tumor, respectively, while eliminating one real MRI sequence from the segmentation procedure.

\end{abstract}

\section{Introduction}

\begin{figure}[!ht]
	\centering
	\includegraphics[width=.95\columnwidth]{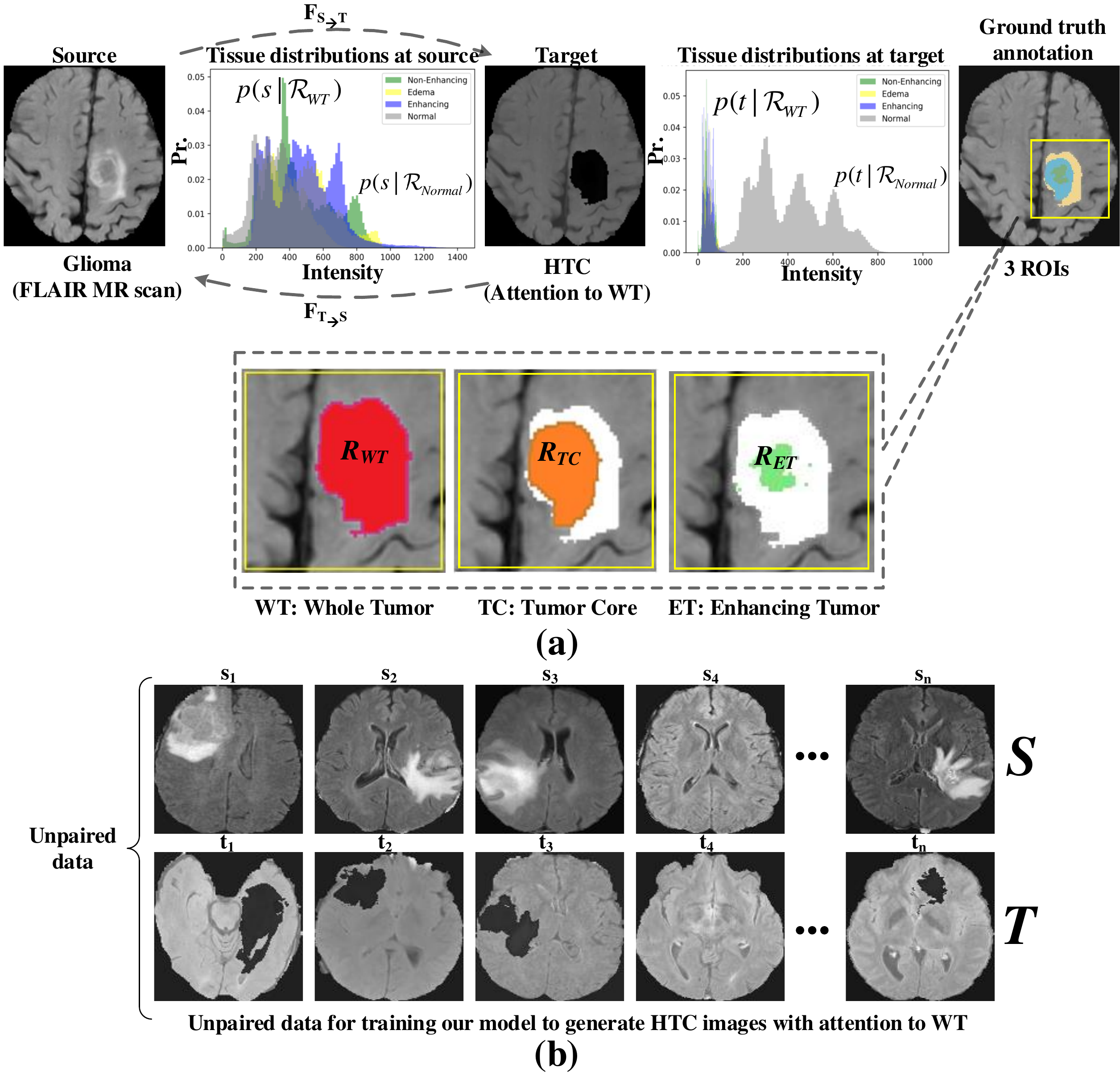} 
	\caption{The MR images show low tissue contrast in the source domain, implying the most challenging for tissue segmentation. (a) A glioma lesion in the FLAIR MR image (left) with its intensity distributions (i.e., non-enhancing, edema, and enhancing) as well as normal tissue distribution. The corresponding HTC target image (middle) with attention to WT is achieved based on the manual labels (right). (b) Unpaired training data, consisting of a source set (first row) $s \sim {p}(s)$ and a target set (second row) $t \sim {{p}}(t)$, with no information provided as to which $s$ matches which $t$.}
	
	\label{fig:overview}
\end{figure}

Among brain tumors, glioma is the most prevalent tumor that begins from the tissue of the brain and can affect the brain function \cite{Elazab}. In brain magnetic resonance (MR) images, the intensity distributions of pixels are largely overlapping in regions of interest (ROIs), therefore leading to low tissue contrast and creating the main challenge for tissue segmentation. In glioma, ROIs exhibit similar levels of intensity in MR images, making tissue segmentation quite challenging. Fig. \ref{fig:overview}(a) demonstrates a brain lesion in the FLAIR MR slice with three overlapping tissues: whole tumor (WT), tumor core (TC), and enhancing tumor (ET). We define target images as a high-contrast domain such that tissues have limited overlapping area, while the source domain has overlapping tissue distributions. Our goal is to increase the intensity contrast between the underlying tissue region and others through image-to-image translation technique based on unpaired training data (Fig. \ref{fig:overview}(b)) to improve segmentation performance.

\begin{figure}[!t]
	\centering
	\includegraphics[width=.95\columnwidth]{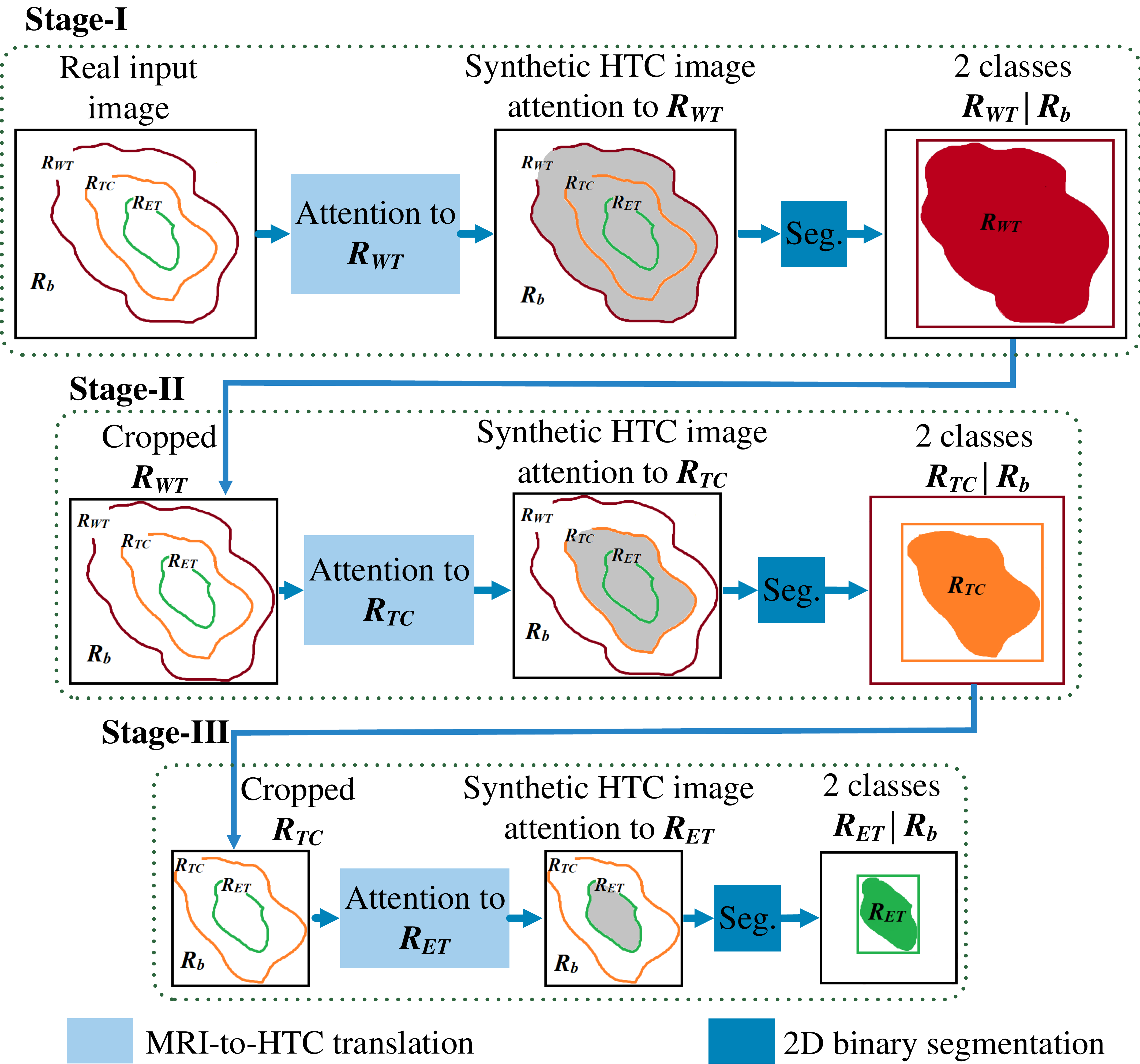} %FigGAN7.pdf
	\caption{Design of the proposed multi-stage structure for segmentation of glioma within three stages. In each step, we first generate synthetic HTC images with the minimum overlapping area for the class-conditional densities between $({\mathcal{R}_{f}})$  and $({{\mathcal{R}}_{b}})$  through MRI-to-HTC block. Then, the synthetic HTC images are applied for segmentation in both end-to-end and two-stage training tactics. Synthetic HTC images deal with ${{\mathcal{R}}_{WT}}$, ${{\mathcal{R}}_{TC}}$, and ${{\mathcal{R}}_{ET}}$ in each stage, sequentially.}
	\label{fig:GAN}
\end{figure}

Image translation aims to learn the mapping between an input image following the source domain distribution to an output image with a defined distribution using a training set of paired \cite{isola2017image} or unpaired images \cite{zhu2017unpaired}. Despite the limitations of the synthetic images in the clinical application, these data have suggested promising results through generative adversarial networks (GANs) \cite{goodfellow2014generative}, including data augmentation \cite{bowles2018gansfer}, image reconstruction \cite{sharma2019missing}, and segmentation \cite{huo2018synseg,Zhang2018,chartsias2017adversarial,Nie2018,wolterink2017deep,zhao2017supervoxel}.
The paired methods require training source images that aligned with the target ones to learn image generation through the forward adversarial loss, while the unpaired approaches frequently employ unaligned images through the CycleGAN structure.

The CycleGAN models have two main components, i.e., a source-to-target and a target-to-source block. Each part consists of a generator \textit{G} and a discriminator \textit{D}. \textit{G} aims to generate a real image from a noise vector and an input image, while \textit{D} is finding the difference between an actual image and the image produced by \textit{G}. 
The key challenges in medical image synthesis either inter-modality (T1-to-T2, FLAIR-to-T1, and others) or cross-modality (MRI-to-CT, PET-to-CT, PET-to-MRI) translation are to predict the structure and fine-grained content of the target modality from the source one \cite{huo2018synseg,Zhang2018,chartsias2017adversarial,Nie2018,wolterink2017deep,zhao2017supervoxel}. The CycleGAN provides effective supervision using cycle consistency between the source inputs and the reconstructed images as well as between the target images and corresponding reconstructed ones.

However, state-of-the-art medical image synthesis methods are restricted by the model's disability to attend a specific tissue. 
In this paper, we propose a multi-stage model to segment only one tissue through a segmentation block following an attention-guided synthesis block in each stage (Fig. \ref{fig:GAN}). Specifically, the synthesis block generates high tissue contrast (HTC) images with attention to the relevant tissue for the segmentation task. In image synthesis block, we have two mappings: MRI-to-HTC and HTC-to-MRI. The former accepts 2D MR slices and generates HTC images, which further fed into the latter to reconstruct the input MR images. In the segmentation block, the HTC images are passed to the convolution layers to produce a binary segmentation map and a bounding box for the next stage.
To provide attention to the specific tissue during synthesis process, two strategies have been used: (1) attachment of the attention block into the CycleGAN, (2) using high contrast image during training phase.
The attention block guides $G$ towards the expected region for translation via an attention map. This trainable map is further employed in $D$ input to filter out irrelevant areas. Regarding the training, we use the ground-truth (GT) labels to form images with the minimum overlapping area between the tissue intensity distribution of the foreground and background in each stage (depicted tissue distribution in Fig. \ref{fig:overview} (a)).  

Furthermore, to produce a more detailed synthesis and consequently more accurate segmentation map, we explore the multi-stage architecture to deal with only one region in each stage. This structure alleviates the artifacts of the synthesized images by decreasing the gap between the source and target domain. 
The attention module effectively learns attention maps to guide the generator attentively select more important regions for generating an HTC image. The generated HTC image closely follows the distribution of the target domain and boosts the segmentation performance significantly.
Besides, our model is based on the CycleGAN framework to leverage the vast quantities of unpaired data sets for training within the same modality. 
The experiments are conducted on multi-modal BraTS 2018 dataset \cite{Menze2015Brain} to segment internal parts of glioma. Specifically, we employ real modalities, i.e., FLAIR, T2, and T1c, to generate synthetic one with attention to the WT, TC, and ET in each stage, respectively. 
The contributions of this paper are summarized as: 

\begin{itemize}
	
	\item  We design a novel framework to increase the contrast among sub-regions of glioma in MR images. Training on high contrast images as well as an unsupervised attention block inside the adversarial network guide our model to pay attention to the particular regions.
	
	\item We propose a multi-stage structure that decreases the gap between the source and target domain to enhance the resolution of synthetic HTC images.
	
	\item We employ HTC MR images in both the end-to-end and two-stage segmentation structure on BraTS dataset to confirm the effectiveness of these images.

\end{itemize}

%The remainder of the paper is organized as follows. In the next Section, we %describe the related works.
%In Section \ref{Method}, the proposed multi-stage attention-guided method %is described in detail.  Section \ref{Experiments} presents the experiments %and results of the proposed method, and finally, Section \ref{conclusion} %concludes this paper.

\section{Related works}
\subsection{Segmentation}
Numerous machine learning \cite{Hatami} and deep learning methods have been introduced to address segmentation problems, especially glioma subregions \cite{Soleymanifard}. Fully convolutional networks (FCNs) \cite{Long2015,ronneberger2015u,Drozdzal2016,Chen2018,jegou2017one} as an extension of convolutional neural networks (CNNs) \cite{he2016deep,huang2017densely} with down-sampling and up-sampling layers have been considered as a benchmark of segmentation.
Replacement of fully connected layers with convolution layers facilitates FCNs to take the global features and provides localization in an end-to-end framework \cite{Long2015}. In U-Net \cite{ronneberger2015u}, authors used U-shaped architecture of FCNs with the skip connection to combine features extracted in the encoder side to the decoder ones. In other work, Drozdzal \textit{et al.} \cite{Drozdzal2016} added the \textit{residual blocks} \cite{he2016deep} to the U-Net framework to improve the segmentation accuracy by reducing the effect of vanishing gradient (Res-U-Net). Chen \textit{et al.} \cite{Chen2018} also extended the fully convolutional version of residual networks (ResNets) \cite{he2016deep} by incorporating the dilation to the main structure.
Jegou \textit{et al.} \cite{jegou2017one} continued the DenseNet \cite{huang2017densely} to fully convolutional DenseNet (FC-DenseNet) without post-processing for segmentation. This architecture leads to implicit in-depth supervision and allows capturing contextual information.

\subsection{Segmentation in Adversarial Framework}
Adversarial methods have been successfully exploited in medical image analysis to address the shortage of large and diverse annotated databases \cite{bowles2018gansfer},  missing/corrupted MR pulse sequences \cite{sharma2019missing}, as well as boost the segmentation performance in typical applications. 
These latter approaches can be categorized as two-stage training techniques \cite{chartsias2017adversarial,wolterink2017deep,zhao2017supervoxel,Nie2018,hamghalam2019brain} and end-to-end methods \cite{huo2018synseg,Zhang2018}. The former considers the synthesis and segmentation as two individual training stages, while the latter incorporates the segmentation loss into the adversarial loss during the training.

Chartsias \textit{et al.} \cite{chartsias2017adversarial} produced synthetic cardiac data from unpaired images coming from different individuals (CT-to-MRI cardiac image) based on CycleGAN. They found that training on both real and synthetic images lead to a statistically significant improvement compared to training on real data.
Wolterink \textit{et al.} \cite{wolterink2017deep} proposed  MRI-to-CT synthesis on pairwise aligned training images of the same patient in the treatment planning of brain tumors. They analyzed paired and unpaired image mapping from 2D brain MR image slices into 2D CT ones. Authors found that the synthetic CT images taken via the model trained with unpaired data seemed more realistic, contained fewer artifacts than those obtained through the model trained with paired data.
Zhao \textit{et al.} \cite{zhao2017supervoxel} introduced a multi-atlas based hybrid approach to
synthesize T1w MR images from CT and CT images from T1w MR images using random forest synthesis framework. This method used a set of random forest regressors within each label for synthesizing intensities on pairs of MR and CT images of the whole head. 
In other works, Nie \textit{et al.} \cite{Nie2018} first applied FCN Model to generate MR from CT image as well as 7T MR from 3T MR images based on the CycleGAN. Next step, they employed synthetic images for the task of semantic segmentation.

In the end-to-end framework, Huo \textit{et al.} \cite{huo2018synseg} integrated the CycleGAN and segmentation into an end-to-end structure to train a segmentation network for both MRI-to-CT and CT-to-MRI without having manual labels in the target modality. In their architecture, called SynSeg-Net, authors demonstrated that end-to-end training achieved better performances compared to the two-stage one for segmentation.
Zhang \textit{et al.} \cite{Zhang2018} presented 3D cross-modality synthesis approach (CT-to-MRI) to segment cardiovascular volumes by adding shape-consistency loss to the CycleGAN framework. They also validated that coupling the generator and segmentor module resulted in better segmentation accuracy than training them exclusively.

\section{Method}
\label{Method}
The proposed framework is composed of $K$ stages, where $K$ denotes the number of labels in input images. Each step consists of two main modules: (1) image synthesis with attention, and (2) segmentation block. The former is learned in an adversarial framework to generate synthetic HTC images with attention to an individual foreground $\mathcal{R}_{f}^{(k)}$, while the latter performs supervised binary segmentation for the foreground and background $\mathcal{R}_{b}^{(k)}$ region. 
The bounding box which calculated from the segmentation map in stage $k$ will be considered for the next step, $k+1$. Fig. \ref{fig:GAN} shows an overview of the proposed structure for segmentation of brain lesion with three regions ($K=3$), including ${\mathcal{R}_{WT}}$, ${\mathcal{R}_{TC}}$, and ${\mathcal{R}_{ET}}$. 
%The main notations are demonstrated on Table \ref{tab:notation}. 
%
This section first describes how the image synthesis block transforms the tissue intensity distribution of the foreground from source to target domain, and then provides details of incorporating the synthetic images into the segmentation framework which is expected to produce more accurate results than using real MR images.

\subsection{HTC Image Synthesis via Attention-GAN (MRI-to-HTC)}

Let MR source image at stage $k$, ${{s}^{(k)}}\in {{S}^{(k)}}$, be the union of foreground, $s_{f}^{(k)}$, and background pixels, $s_{b}^{(k)}$, in the source domain as:
\begin{equation}\label{eq000}
\begin{aligned}
s=[{{s}_{f}}\sim p(s|{{\mathcal{R}}_{f}})]\cup [{{s}_{b}}\sim p(s|{{\mathcal{R}}_{b}})]
\end{aligned}
\end{equation}
we omit the superscript $k$ for simplicity. Similarly, in the target domain, we have HTC image, ${{t}^{(k)}}\in {{T}^{(k)}}$ as:
\begin{equation}\label{eq00}
\begin{aligned}
t=[{{t}_{f}}\sim p(t|{{\mathcal{R}}_{f}})]\cup [{{t}_{b}}\sim p(t|{{\mathcal{R}}_{b}})]
\end{aligned}
\end{equation}
where $p(s|\mathcal{R})$ and $p(t|\mathcal{R})$ are the class-conditional distributions of tissue in the source and target domain, respectively. We also assume that the distribution of the foreground and background have a little overlap in the target space.

Our goal in each stage is to estimate a mapping function, ${{F}_{{S}\to{T}}}$: MRI-to-HTC, from a source domain ${S}$ (MRI image) to the target domain $T$ (HTC image) based on independently sampled data instance, such that the distribution of the mapped samples, $s'$, matches the probability distribution $p(t)$ of the target. For the cycle consistency, a domain inverse mapping, ${{F}_{{T}\to{S}}}$: HTC-to-MRI, also generates the reconstructed images, $s''$, to match closely to the input image $s\approx s''$.

\subsubsection{Attention Block}

In our mapping, we need to generate HTC images that provide maximum segmentation accuracy in $\mathcal{R}_{f}$. To this end, we first need to locate the $\mathcal{R}_{f}$ to translate in each image and then apply the translation to that region.
Specifically, we achieve this by adding two attention networks $\mathcal{{A}_{S}}$ and $\mathcal{A}_{T}$, which select areas to translate by maximizing the probability that the discriminator makes a mistake in the source and target domain, respectively.
The attention block is an FCN network consists of convolution, deconvolution, and the ResNet \cite{he2016deep} unit, followed by the soft-max layer. For each input image, it produces a per-pixel attention map with the same size of the input image indicating the importance of the spatial information. Mainly, after feeding the input image to the generator, we employ the attention mask to the generated image using an element-wise product ($\odot$), and then add the background using the inverse of the mask applied to the input image.  

\begin{figure}[!t]
	\centering
	\includegraphics[width=.95\columnwidth]{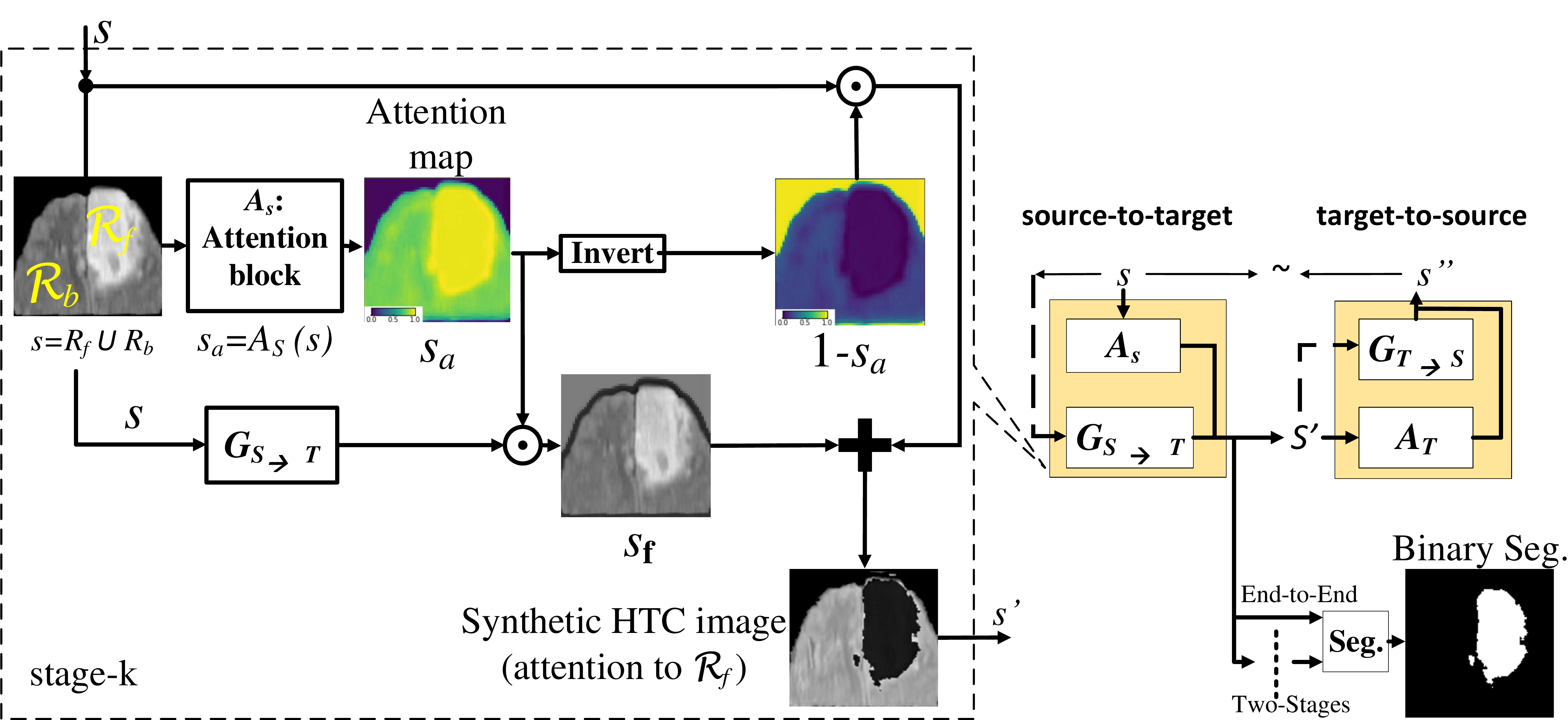} %FigGAN7.pdf  Attention3
	\caption{Image synthesis with attention to WT at stage-I. MRI and HTC images are considered as source and target.}
	\label{fig:ATTENTION}
\end{figure}

As shown in Fig. \ref{fig:ATTENTION}, $s$ is split into two parts: the first part is fed to the source attention block, $\mathcal{A}_{S}$, to create the attention map, $s_{a}=\mathcal{A}_{S}(s)$, while the second part is considered as an input of the generator $G_{S\to T}$ to highlight the foreground region. To eliminate the background region, $s_{a}$ is element-wisely multiplied by $G_{S\to T}(s)$ to make masked image as: $s_{f}=s_{a}\odot G_{S\to T}(s)$. Finally, the synthetic HTC image can be calculated as:
\begin{equation}\label{eq0}
\begin{aligned}
{s}'=s_{a}\odot {{G}_{{{S}}\to {{T}}}}({{s}})+(1-s_{a})\odot {{s}}
\end{aligned}
\end{equation}
where ${s}'$ is passed to the segmentation block to segments the $\mathcal{R}_{f}$  and fed to the domain inverse mapping for the reconstruction. Likewise, we have:
\begin{equation}\label{eq01}
\begin{aligned}
s''=t_{a}\odot {{G}_{{{T}}\to {{S}}}}({{s'}})+(1-t_{a})\odot {{s'}}
\end{aligned}
\end{equation} 
where $t_{a}=\mathcal{A}_{T}(s')$ is the attention map in target domain. 
\subsubsection{Training Procedure}
The training of MRI-to-HTC network requires a discriminator $D_{T}$ to discern the translated outputs from the real HTC images $t$.
Likewise, the discriminator at the source domain, $D_{S}$, encourages HTC-to-MRI network to translate $t$ into source domain indistinguishable from the source domain. We train both discriminators such that they only rate attended regions.
Particularly, instead of employing an entire image as the input, we first filter both generated and real image via an element-wise multiplication with the attention map at source and target domain. Then the filtered images are fed into the discriminator for evaluation. According to \cite{mejjati2018unsupervised}, to avoid the mode collapse, we train the network on whole images ($s$ and $t$) for 25 epochs and then switch to masked ones ($s\odot s_{a}$  and $t\odot t_{a}$), when the attention blocks $\mathcal{A}_{S}$ and $\mathcal{A}_{T}$ have trained moderately.

According to Equations \ref{eq0} and \ref{eq01}, as long as $\mathcal{A}_{S}$ and $\mathcal{A}_{T}$ attend to the background regions, the generated images will preserve their input domain classes. Thus, the discriminators can simply detect the images as fake ones.  To be thriving in two-player minimax game, $\mathcal{A}_{S}$ and $\mathcal{A}_{T}$ have to concentrate on the objects or regions that the corresponding discriminator thinks are the most descriptive within its domain (i.e., the foreground). Finally, the network finds an equilibrium between the generator, attention map, and discriminator to produce realistic images.
\begin{figure}[!t]
	\centering
	\includegraphics[width=.95\columnwidth]{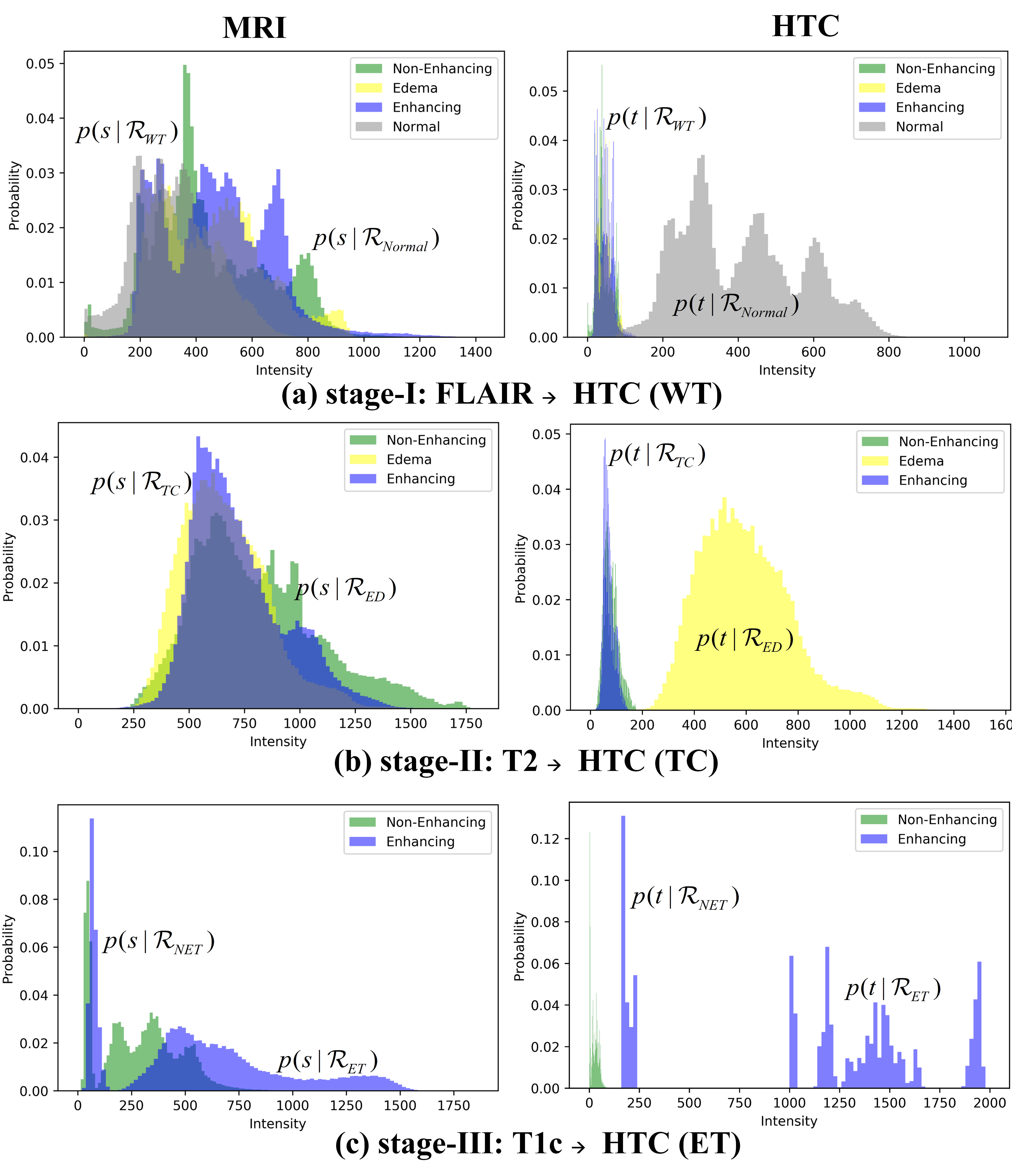} %FigGAN7.pdf
	\caption{Tissue distributions of glioma at the source (left) and target (right) domain. (a) FLAIR MR images are used in the first stage to produce synthetic HTC images with attention to $\mathcal{R}_{WT}$. (b) At the second stage, T2 images are cropped according to the bounding box at the first stage and employed to increase tissue contrast between $\mathcal{R}_{TC}$ and Edema $\mathcal{R}_{ED}$. (c) The third stage is dedicated to enhancing $\mathcal{R}_{ET}$ and non-enhancing tumor $\mathcal{R}_{NET}$ from T1c MR images.}
	\label{fig:histogram_lesion}
\end{figure}
\subsubsection{Preparing Target HTC Images}\label{Preparing target HTC images}
The assumption of non-overlapping tissue distribution in the target domain can be achieved through the GT labels. We change the class-conditional distributions of $p(t|{{\mathcal{R}}_{f}})$ and $p(t|{{\mathcal{R}}_{b}})$ according to the manual labels as depicted in Fig. \ref{fig:histogram_lesion}. We minimize the inter-class variance while maximizing the intra-class distance between $p(t|{{\mathcal{R}}_{f}})$ and $p(t|{{\mathcal{R}}_{b}})$ in the target space. Value of the mean and variance in the target domain are considered as a hyperparameter. Choosing an appropriate amount will produce maximum segmentation accuracy. However, there is a trade-off between class overlap at large values and visual artifacts at small ones. Particularly, low values for variance will generate much sharper results but introduces
visual artifacts, which leads to a decline in segmentation performance. Fig. \ref{fig:histogram_lesion} (left column) demonstrates the considerable class overlap between the distributions of the foreground and background tissue in the source domain on the BraTS dataset. We use FLAIR, T2, and T1c sequence to segment WT, TC, and ET in each stage, respectively. Fig. \ref{fig:histogram_lesion} (right column) shows distributions of the corresponding tissues in the defined target domain.

%\caption{Distribution of foreground and background of brain lesions in source (left column) and target (right %column) domain. The network takes as input brain lesion MR images that are cropped to the ROI (with %resolution ${w}_{k}\times {w}_{k}$): (a) to the whole brain, (b) to the whole tumor and (c) the enhancing %tumor.}

\subsection{Segmentation Block}
Segmentation block provides feedback for the image synthesis one during the training in the case of end-to-end strategy.
We apply the 2D  binary segmentation structure with the weighted cross-entropy loss, ${{\mathcal{L}}_{seg}}$, to handle the class imbalance, especially in the first stage.
Specifically, FC-DenseNet comprises the Dense blocks (batch normalization (BN), ReLU, $3\times3$ convolution, and Dropout), the Transition down blocks (BN, ReLU, $1\times1$ convolution, Dropout, and $2\times2$ Max Pooling), and the Transition up block ( $3\times3$ Transposed convolution with stride of 2). We also consider non-overlapping max pooling and Dropout with $p=0.2$. Each Dense block contains four layers of convolution which each layer calculates 12 feature maps. These features are sequentially concatenated to build 48 feature maps at the output of Dense block.
In the training phase, the bounding boxes are automatically generated based on the GT, whereas, in the testing phase, the bounding boxes are obtained based on the binary segmentation results of the preceding stage.

\subsection{Loss Functions}

In addition to the segmentation loss, ${\mathcal{L}}_{Seg.}$, there are four loss functions to generate HTC images in each stage. The adversarial loss to take advantage of GAN networks at the source, $\mathcal{L}_{adv}^{s}$, and target domain, $\mathcal{L}_{adv}^{t}$, as:
\begin{equation}\label{eq1}
\begin{aligned}
\mathcal{L}_{adv}^{s}({{F}_{S\to T}},{{A}_{S}},{{D}_{T}})={{\mathbb{E}}_{t\sim {{p}}(t)}}[\log ({{D}_{T}}(t))]+\\{{\mathbb{E}}_{s\sim {{p}}(s)}}[\log (1-{{D}_{T}}({s}'))]
\end{aligned}
\end{equation}
%%%
\begin{equation}\label{eq2}
\begin{aligned}
\mathcal{L}_{adv}^{t}({{F}_{T\to S}},{{A}_{T}},{{D}_{S}})={{\mathbb{E}}_{s\sim {{p}}(s)}}[\log ({{D}_{S}}(s))]+\\{{\mathbb{E}}_{t\sim {{p}}(t)}}[\log (1-{{D}_{S}}({t}'))]
\end{aligned}
\end{equation}
%%%%
%%%%
Meanwhile, and similarly to CycleGAN, we add a cycle-consistency loss to the adversarial ones by enforcing a one-to-one mapping between true image, $s$, and cycle reconstructed ones, ${s}''$, as a forward cycle consistency loss:
%%%%%
\begin{equation}\label{eq3}
\begin{aligned}
\mathcal{L}_{cyc}^{s}(s,{s}'')={{\left\| s-{s}'' \right\|}_{1}}
\end{aligned}
\end{equation}
where ${s}''={{F}_{T\to S}}({{F}_{S\to T}}(s))$. In the backward path, we also have the backward cycle consistency loss as:
%%%
%%%
\begin{equation}\label{eq4}
\begin{aligned}
\mathcal{L}_{cyc}^{t}(t,{t}'')={{\left\| t-{t}'' \right\|}_{1}}
\end{aligned}
\end{equation}
where ${t}''={{F}_{S\to T}}({{F}_{T\to S}}(t))$. Finally, we combine the defined loss functions with different weights. The final objective for the image synthesis, ${\mathcal{L}}_{synth.}$, is:
%%
%%%
\begin{equation}\label{eq5}
\begin{aligned}
{{\mathcal{L}}_{synth.}}={{\lambda }_{1}}.\mathcal{L}_{adv}^{s}({{F}_{S\to T}},{{A}_{s}},{{D}_{T}})+\\{{\lambda }_{2}}.\mathcal{L}_{cyc}^{s}({{F}_{S\to T}},{{F}_{T\to S}},S)+\\{{\lambda }_{3}}.\mathcal{L}_{adv}^{t}({{F}_{T\to S}},{{A}_{T}},{{D}_{S}})+\\{{\lambda }_{4}}.\mathcal{L}_{cyc}^{t}({{F}_{T\to S}},{{F}_{S\to T}},T)
\end{aligned}
\end{equation}
where ${\lambda }_{1}$, ${\lambda }_{2}$, ${\lambda }_{3}$, and ${\lambda }_{4}$ are the scalar hyper-parameters to regularize the loss functions.

In the two-stage training strategy, we first minimize ${\mathcal{L}}_{synth.}$ to generate HTC images with attention to the specific region. Then, we optimize ${\mathcal{L}}_{Seg.}$ with fixed synthesis loss, as two independent training steps. While, in the case of end-to-end, we optimize ${\mathcal{L}}_{total}$ which incorporates the segmentation loss into the adversarial one during training as:
\begin{equation}\label{eq5}
\begin{aligned}
{\mathcal{L}}_{total}={\lambda }_{5}{\mathcal{L}}_{Seg.} + {\mathcal{L}}_{Synth.} 
\end{aligned}
\end{equation}
where ${\lambda }_{5}$ balances the effect of ${\mathcal{L}}_{Seg.}$ to equip our HTC synthesis model with the segmentation feedback.

\section{Experiments and Results}\label{Experiments}

We conduct several experiments on BraTS 2018 dataset to demonstrate the effectiveness of the proposed methods on the synthesizing HTC images and the segmentation task. Each sequence has been normalized separately by subtracting the mean and dividing by the standard deviation of the brain region, and the non-brain area is set to zero. Furthermore, all networks are trained for 180 epochs with Adam learning rate of 0.0001. Our implementation is developed employing TensorFlow on an NVIDIA TITAN X GPU with 12G of RAM. 

%Larger patch sizes across the brain are preferred as they can provide brain structures in the training process, %while smaller patches employ only local information and allow extending the batch size in the optimization %process. We choose patches of size $64\times64\times64$ with the central voxel at brain tissue.

\subsubsection{Dataset}

The performance of the proposed method is evaluated on publicly available BraTS 2018 dataset \cite{Menze2015Brain,bakas2017advancing,Bakas2018_Identifying}, gathered from various scanners with an in-plane matrix size of $240\times240\times155$. Four MR sequences are available for each patient consist of FLAIR, T1, T1c,  and T2. Evaluation is performed for three ROIs, including the WT (all internal parts of tumor), TC (enhancing and non-enhancing), and ET.
Around 2K axial slices are randomly selected and center-cropped to $128\times128$ for training, such that each slice has non-zero value on at least half of the pixels. 
Specifically, the first stage, 2D FLAIR MR images are used to generate synthetic HTC one in our cyclic framework with attention to WT as: FLAIR$\leftrightarrow$$\textrm{FLAIR}^{'}$. Then, we segment $\textrm{FLAIR}^{'}$with the end-to-end (FLAIR$\leftrightarrow$$\textrm{FLAIR}^{'}$$\rightarrow$${\mathcal{R}_{WT}}$) as well as the two-stage (FLAIR$\leftrightarrow$$\textrm{FLAIR}^{'}$, $\textrm{FLAIR}^{'}$$\rightarrow$${\mathcal{R}_{WT}}$) approach. Accordingly, for the segmentation of TC, we extract T2 patches to $96\times96$ from the corresponding slices. Thus, we have: T2$\leftrightarrow$$\textrm{T2}^{'}$$\rightarrow$${\mathcal{R}_{TC}}$ and T2$\leftrightarrow$$\textrm{T2}^{'}$, $\textrm{T2}^{'}$$\rightarrow$${\mathcal{R}_{TC}}$ for the end-to-end and two-stage, respectively. In the last stage, segmentation of ET, we apply T1c patches with a size of $64\times64$ to generate the synthetic HTC images and predict pixel labels of ET (T1c$\leftrightarrow$$\textrm{T1c}^{'}$$\rightarrow$${\mathcal{R}_{ET}}$ and T1c$\leftrightarrow$$\textrm{T1c}^{'}$, $\textrm{T1c}^{'}$$\rightarrow$${\mathcal{R}_{ET}}$).

\subsubsection{Evaluation of the Synthetic HTC MR Images}

\begin{figure}[!t]
	\centering
	\includegraphics[width=.95\columnwidth]{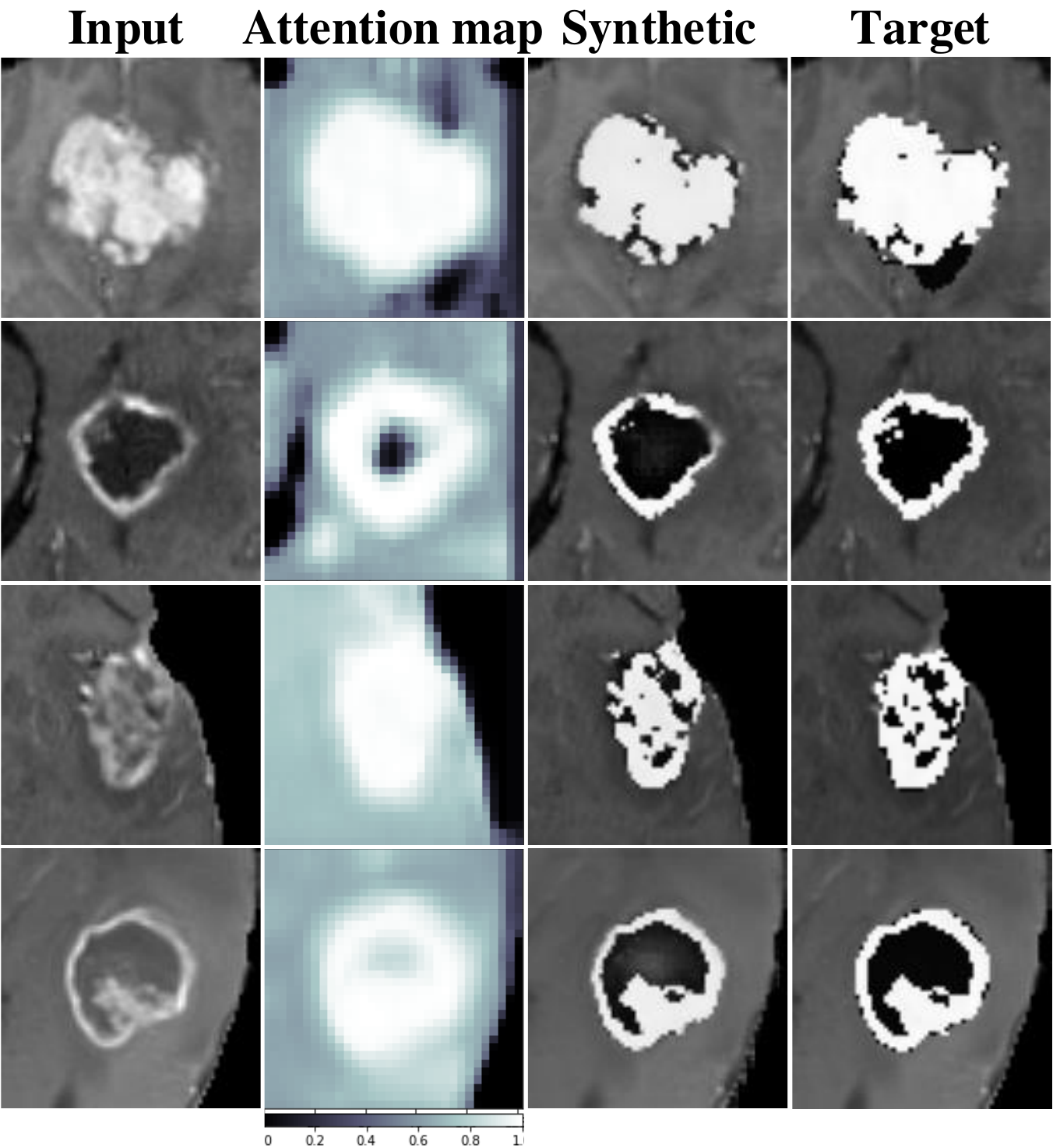}   %%%ET.pdf}
	\caption{Examples of MRI-to-HTC translation with attention to ET on BraTS dataset. From left to right: input image, attention map, synthetic HTC image, and target image.}
	\label{fig:ET}
\end{figure}

To evaluate the synthetic HTC MR images, we calculate the Kolmogorov-Smirnov (K-S) statistic on the target domain to estimate the goodness-of-fit between the intensity distribution of the synthetic HTC and real HTC images for each class label. Table \ref{tab:KS-test} lists the consequences of the K-S test for the WT, TC, ET, and Normal of the brain tumor for various loss weight values. Note that the segmentation block is bypassed (${\lambda }_{5}=0$) to assess the quality of synthetic HTC images.
\begin{table}[t]
	\centering
	\caption{K-S test results between the synthetic and target HTC images applying different loss weights.} 
	\label{tab:KS-test}
	\begin{tabular}{ccccccc}
		\toprule[1pt] \midrule[0.3pt]  	
		\multicolumn{2}{c}{\textbf{Loss weights}} && \multicolumn{4}{c}{\textbf{Brain lesions}} \\
		\cmidrule[\heavyrulewidth]{1-2}  \cmidrule[\heavyrulewidth]{4-7}
		${\lambda }_{1}, {\lambda }_{3}$ & ${\lambda }_{2}$, ${\lambda }_{4}$  && WT & TC & ET & Normal \\
		\midrule 
		
		1      & 10    && \textbf{0.31}& \textbf{0.30} &\textbf{0.34} & \textbf{0.14}  \\ 
		10     & 1     && 0.28& 0.26 &0.22 & 0.10 \\ 
		1      & 100    && 0.30& 0.27 &0.32 & 0.06 \\ 
		1      & 1     && 0.12& 0.18 &0.12 & 0.13 \\ 
		\midrule[0.3pt]\bottomrule[1pt]
	\end{tabular} 
\end{table} 
We further appraise the quality of synthetic HTC images on each stage using peak signal-to-noise ratio (PSNR) and structural similarity index metric (SSIM) in Table \ref{tab:quality_assessment}. In these experiments, the loss weights are  considered as ${\lambda }_{1}, {\lambda }_{3} = 1$ and ${\lambda}_{2}$, ${\lambda }_{4} = 10$. 
\begin{table}[!t]
	\centering
	\caption{Quality evaluation of the synthetic HTC images in our multi-stage framework.} 
	
	\begin{tabular}{cccc}
		\toprule[1pt] \midrule[0.3pt]
		\rule[-1ex]{0pt}{2.5ex} \textbf{MRI-to-HTC (attention)}	& SSIM & PSNR\\ 
		\midrule 
		
		\rule[-1ex]{0pt}{2.5ex}	$\textrm{FLAIR}\leftrightarrow\textrm{FLAIR}^{'}(WT)$ & 0.6132 & 17.37 \\ 
		\rule[-1ex]{0pt}{2.5ex} $\textrm{T2}\leftrightarrow\textrm{T2}^{'}(TC)$       & 0.6284 & 18.32 \\ 
		\rule[-1ex]{0pt}{2.5ex} $\textrm{T1}\leftrightarrow\textrm{T1c}^{'}(ET)$      & 0.6449 & 19.87 \\
		
		\midrule[0.3pt]\bottomrule[1pt]
	\end{tabular} 
	\label{tab:quality_assessment}
\end{table}
Moreover, Fig. \ref{fig:ET} shows examples of synthetic HTC images with attention to ET at stage III. The first column presents the real T1c MR patches in the source domain, the second column displays the attention maps, and the third one shows the corresponding synthetic HTC patches, and the last column depicts the real HTC images in the target domain.

\subsubsection{Ablation Analysis}
We measure PSNR and SSIM as similarity metrics between the synthetic HTC and target images. In Table \ref{tab:ablation_study}, we first employ the plain CycleGAN \cite{zhu2017unpaired} to generate HTC images with attention to four regions. Then, we judge the model with only one attention block in either the source (CycleGAN+$\mathcal{A}_{S}$) or the target domain (CycleGAN+$\mathcal{A}_{T}$). Finally, we repeat the experiment to consider only one region (ET) to assess our multi-stage  MRI-to-HTC structure with attention blocks. 
\begin{table}[!t]
	\centering
	\caption{Ablation study of multi-stage MRI-to-HTC model.} 
	
	\begin{tabular}{cccc}
		\toprule[1pt] \midrule[0.3pt]
		\rule[-1ex]{0pt}{2.5ex} \textbf{Model}	& \textbf{SSIM}   & \textbf{PSNR}  \\ 
		\midrule 
		
		\rule[-1ex]{0pt}{2.5ex} CycleGAN                								& 0.6072 & 17.64 \\ 
		\rule[-1ex]{0pt}{2.5ex} CycleGAN+$\mathcal{A}_{S}$							& 0.6384 & 18.21 \\ 
		\rule[-1ex]{0pt}{2.5ex} CycleGAN+$\mathcal{A}_{t}$    		  				& 0.6387 & 18.29 \\ 
		\rule[-1ex]{0pt}{2.5ex} CycleGAN+$\mathcal{A}_{S}$+$\mathcal{A}_{t}$   		& 0.6647 & 18.86 \\ 
		\rule[-1ex]{0pt}{2.5ex} CycleGAN+$\mathcal{A}_{S}$+$\mathcal{A}_{t}$+multi-stage	& \textbf{0.6849} & \textbf{19.87} \\ 
	
		\midrule[0.3pt]\bottomrule[1pt]
	\end{tabular} 
	\label{tab:ablation_study}
\end{table}
\subsubsection{Comparisons with Other Synthetic Segmentation Methods}
We compare our method with recently proposed approaches \cite{huo2018synseg} and \cite{chartsias2017adversarial}, which employed synthetic images for segmentation in the end-to-end and two-stage manner, respectively. The former combines the segmentation loss with the adversarial one during training, while the latter individually trains the image synthesis and segmentation block. In Table \ref{tab:Brain_end_to_end}, we measure the segmentation accuracy for WT, TC, and ET via the 4-fold cross-validation and observe that the proposed end-to-end method with the attention block achieves the highest accuracy compared to others. Table \ref{tab:Brain_end_to_end} also demonstrates the advantage of end-to-end training over the two-stage one in terms of accuracy. Note that we need roughly 27 ms to generate synthetic HTC image (2D) in the two-stage framework during the inference time. 
\begin{table*}[!htbp]
	\centering
	\caption{Segmentation accuracy for WT, TC, and ET of brain lesion in MR images via cross-validation.} 
	\label{tab:Brain_end_to_end}
	\begin{tabular}{@{\extracolsep{\fill}}cccccccc}
		\toprule[1pt] \midrule[0.3pt]
		\rule[-1ex]{0pt}{2.5ex}
		\multirow{2}{*}{\textbf{Dice}} &&   &  \multicolumn{2}{c}{\textbf{CycleGAN + Segmentation}} && \multicolumn{2}{c}{\textbf{Proposed}}  \\
		\cmidrule[\heavyrulewidth]{4-5} 
		\cmidrule[\heavyrulewidth]{7-8}
		\textbf{Mean ($\pm$Std.(\%))}&& &\textbf{End-to-End}  & \textbf{Two-Stage}  && \textbf{End-to-End}  & \textbf{Two-Stage} \\
		\midrule
		WT && &	0.9231 (0.24) &	0.8804 (0.37) &&	\textbf{0.9508 (0.58)}&0.9304 (0.42)\\
		%\midrule
		TC &&  &	0.9016 (0.11) &	0.8652 (0.13) &&\textbf{0.9304 (0.51)}  &	0.9061 (0.17) \\   %82.73 for 10 and 85.36 for 100
		%82.39 for 15
		%	\midrule
		ET  && 	 &	0.8699 (0.14) &0.8370 (0.18) &&	\textbf{0.8891 (0.13)} &	0.8612 (0.11) \\	
		%\midrule 
		\midrule[0.3pt]\bottomrule[1pt]
	\end{tabular}
\end{table*}
\begin{table*}
	\centering
	\caption{Dice scores and HD95 with and without the proposed synthetic HTC images on BraTS'18 Validation datasets.\\ FLAIR* indicates the synthetic HTC MR images by attention to non-enhancing, edema, enhancing, and normal regions.} 
	\label{tab:EnhGAN_cmp_others}
	\begin{tabular}{@{\extracolsep{\fill}}c|c|ccccccc}
		\toprule[1pt] 
		\rule[-1ex]{0pt}{2.5ex}
		\multirow{3}{*}{\textbf{Method}} &\multirow{3}{*}{\textbf{Modality concatenation}}& \multicolumn{3}{c}{\textbf{Dice}} && \multicolumn{3}{c}{\textbf{HD95 (mm)}  }  \\
		\cmidrule[\heavyrulewidth]{3-5} 
		\cmidrule[\heavyrulewidth]{7-9}
		&&\textbf{EN} &\textbf{WT} & \textbf{TC} && \textbf{EN} &\textbf{WT} & \textbf{TC}\\
		
		\midrule 
		\multirow{2}{*}{U-Net  } &FLAIR, T1, T1c, T2 & 0.7874          &	0.8913          &	0.8402          &&	4.15&5.61 & \textbf{7.71}\\
		%\midrule 
		&  FLAIR, FLAIR$\leftrightarrow$$\textrm{FLAIR}^{*}$,T1c,T2          & \textbf{0.7921} &	\textbf{0.8998} &	\textbf{0.8462} &&	\textbf{3.94} &	\textbf{5.24} &	8.02\\
		
		\midrule 
		\multirow{2}{*}{Res-U-Net }  &FLAIR,T1,T1c,T2     &	0.7891         &	0.8951 &	0.8413  &&	4.03&	5.5l&	8.66 \\	
		%\midrule 
		
		&FLAIR, FLAIR$\leftrightarrow$$\textrm{FLAIR}^{*}$,T1c,T2            & \textbf{0.7944} &\textbf{0.9033} &	\textbf{0.8475}&&	4.03&	\textbf{5.02}&\textbf{6.31}\\
		
		\midrule 
		
		\multirow{2}{*}{FC-DenseNet }  &FLAIR,T1,T1c,T2   &	0.7890         &	0.8965 &	0.8439  &&	\textbf{4.05} &	5.41 &	7.95\\	
		%\midrule 
		
		&FLAIR, FLAIR$\leftrightarrow$$\textrm{FLAIR}^{*}$,T1c,T2 & \textbf{0.7943}	&\textbf{0.9041}&	\textbf{0.8498}&&	4.33&	\textbf{4.95}&\textbf{7.75}\\

		\midrule[0.1pt]
	\end{tabular}
\end{table*}
\subsubsection{Synthetic HTC Volumes in 3D Multi-Modal Segmentation Framework}
We evaluate the effect of synthetic HTC images in the 3D multi-modal segmentation framework based on the two-stage training approach. To this end, we substitute T1 MR volume for the corresponding FLAIR$\leftrightarrow$$\textrm{FLAIR}^{*}$ sequence, while increasing contrast among the non-enhancing, edema, enhancing, and normal regions.
We experiment with three state-of-the-art segmentation models, including U-Net \cite{ronneberger2015u}, Res-U-Net \cite{Drozdzal2016}, and FC-DenseNet \cite{jegou2017one}. To have a fair comparison, we perform experiments using four sequences in both cases, i.e., FLAIR, T1, T1c, and T2 for the real segmentation as well as FLAIR, FLAIR*, T1c, T2 for the synthetic one. Since T1 modality has less information regarding glioma compared to other sequences, we eliminate T1 in our experiments. 
Table \ref{tab:EnhGAN_cmp_others} presents Dice and modified Hausdorff distance (HD95) on BraTS'18 validation set (Leaderboard), reported by the CBICA image processing online portal. Segmentation with HTC sequences improves Dice scores in three clinically important sub-regions, including WT (0.8$\%$), TC (0.6$\%$), and ET (0.5$\%$). We also achieve averagely 0.4 mm improvement in WT in terms of HD95. However, we need approximately 4.2s to generate each FLAIR* volume form real FLAIR.
%
%
%
%In both training strategies, the corresponding real scans can be concatenated with synthetic ones to train the %model. Here, we only use one modality for evaluation.
\section{Discussion and Conclusion}
\label{conclusion}
We have shown that a deep neural network can be trained on the unpaired dataset to synthesize an HTC image from an MR image. Our proposed supervised model modify the class-conditional distributions of ROIs for the segmentation task in each stage based on the GAN model, which is equipped with attention mechanisms to alter only relevant regions in the input image. We validate our approach on the sub-regions of glioma in multi-modal MR scans of BraTS 2018 dataset. The results of the K-S test confirm that proposed MRI-to-HTC can modify the distributions of WT, TC, and ET in the FLAIR, T2, and T1c MR images, respectively.
The experiments over three segmentation baselines indicate that incorporating the synthetic HTC images with other modalities, i.e., FLAIR, T1c, and T2, improves Dice score and HD95 on BraTS 2018 Leaderboard while eliminating the T1 MR sequence from the segmentation procedure. Although the proposed MRI-to-HTC can achieve promising results, it still has a limitation on defining the mean and standard deviation of the class-conditional distribution in the HTC target images. Small standard deviation values generate much sharper results but introduce visual artifacts in the synthetic images, which reduce the segmentation accuracy.
As a direction for future works, one can develop a framework to tackle corrupted or missing MR volume, that appears during scanning in the acquisition setting. Towards this end, the synthetic HTC volume can be replaced with the corrupted one to complement the information presented by the missing sequence for automated systems. 
%
%\clearpage
\bibliography{mybib}

\begin{thebibliography}{}

\bibitem[\protect\citeauthoryear{Bakas \bgroup et al\mbox.\egroup
  }{2017}]{bakas2017advancing}
Bakas, S.; Akbari, H.; Sotiras, A.; et~al.
\newblock 2017.
\newblock Advancing the cancer genome atlas glioma {MRI} collections with
  expert segmentation labels and radiomic features.
\newblock {\em Scientific Data} 4:170117.

\bibitem[\protect\citeauthoryear{Bakas \bgroup et al\mbox.\egroup
  }{2018}]{Bakas2018_Identifying}
Bakas, S.; Reyes, M.; Jakab, A.; et~al.
\newblock 2018.
\newblock Identifying the best machine learning algorithms for brain tumor
  segmentation, progression assessment, and overall survival prediction in the
  {BRATS} challenge.
\newblock {\em arXiv preprint arXiv:1811.02629}.

\bibitem[\protect\citeauthoryear{Bowles \bgroup et al\mbox.\egroup
  }{2018}]{bowles2018gansfer}
Bowles, C.; Gunn, R.; Hammers, A.; et~al.
\newblock 2018.
\newblock {GANsfer} learning: Combining labelled and unlabelled data for {GAN}
  based data augmentation.
\newblock {\em arXiv preprint arXiv:1811.10669}.

\bibitem[\protect\citeauthoryear{Chartsias \bgroup et al\mbox.\egroup
  }{2017}]{chartsias2017adversarial}
Chartsias, A.; Joyce, T.; Dharmakumar, R.; et~al.
\newblock 2017.
\newblock Adversarial image synthesis for unpaired multi-modal cardiac data.
\newblock In {\em International Workshop on Simulation and Synthesis in Medical
  Imaging},  3--13.

\bibitem[\protect\citeauthoryear{Chen \bgroup et al\mbox.\egroup
  }{2018}]{Chen2018}
Chen, L.; Papandreou, G.; Kokkinos, I.; et~al.
\newblock 2018.
\newblock Deeplab: Semantic image segmentation with deep convolutional nets,
  atrous convolution, and fully connected {CRFs}.
\newblock {\em IEEE Transactions on Pattern Analysis and Machine Intelligence}
  40(4):834--848.

\bibitem[\protect\citeauthoryear{Drozdzal \bgroup et al\mbox.\egroup
  }{2016}]{Drozdzal2016}
Drozdzal, M.; Vorontsov, E.; Chartrand, G.; Kadoury, S.; and Pal, C.
\newblock 2016.
\newblock The importance of skip connections in biomedical image segmentation.
\newblock In {\em Deep Learning and Data Labeling for Medical Applications}.
  Springer.
\newblock  179--187.

\bibitem[\protect\citeauthoryear{{Elazab} \bgroup et al\mbox.\egroup
  }{2018}]{Elazab}
{Elazab}, A.; {Abdulazeem}, Y.~M.; {Anter}, A.~M.; {Hu}, Q.; {Wang}, T.; and
  {Lei}, B.
\newblock 2018.
\newblock Macroscopic cerebral tumor growth modeling from medical images: A
  review.
\newblock {\em IEEE Access} 6:30663--30679.

\bibitem[\protect\citeauthoryear{Goodfellow \bgroup et al\mbox.\egroup
  }{2014}]{goodfellow2014generative}
Goodfellow, I.; Pouget-Abadie, J.; Mirza, M.; et~al.
\newblock 2014.
\newblock Generative adversarial nets.
\newblock In {\em Advances in Neural Information Processing Systems},
  2672--2680.

\bibitem[\protect\citeauthoryear{Hamghalam, Lei, and
  Wang}{2019}]{hamghalam2019brain}
Hamghalam, M.; Lei, B.; and Wang, T.
\newblock 2019.
\newblock Brain tumor synthetic segmentation in {3D} multimodal {MRI} scans.
\newblock {\em arXiv preprint arXiv:1909.13640}.

\bibitem[\protect\citeauthoryear{{Hatami} \bgroup et al\mbox.\egroup
  }{2019}]{Hatami}
{Hatami}, T.; {Hamghalam}, M.; {Reyhani-Galangashi}, O.; and {Mirzakuchaki}, S.
\newblock 2019.
\newblock A machine learning approach to brain tumors segmentation using
  adaptive random forest algorithm.
\newblock In {\em 2019 5th Conference on Knowledge Based Engineering and
  Innovation (KBEI)},  076--082.

\bibitem[\protect\citeauthoryear{{He} \bgroup et al\mbox.\egroup
  }{2016}]{he2016deep}
{He}, K.; {Zhang}, X.; {Ren}, S.; et~al.
\newblock 2016.
\newblock Deep residual learning for image recognition.
\newblock In {\em IEEE Conference on Computer Vision and Pattern Recognition},
  770--778.

\bibitem[\protect\citeauthoryear{{Huang} \bgroup et al\mbox.\egroup
  }{2017}]{huang2017densely}
{Huang}, G.; {Liu}, Z.; v.~d. {Maaten}, L.; et~al.
\newblock 2017.
\newblock Densely connected convolutional networks.
\newblock In {\em IEEE Conference on Computer Vision and Pattern Recognition},
  2261--2269.

\bibitem[\protect\citeauthoryear{{Huo}, {Xu}, and {Moon}}{2019}]{huo2018synseg}
{Huo}, Y.; {Xu}, Z.; and {Moon}, H.
\newblock 2019.
\newblock Synseg-net: Synthetic segmentation without target modality ground
  truth.
\newblock {\em IEEE Transactions on Medical Imaging} 38(4):1016--1025.

\bibitem[\protect\citeauthoryear{{Isola} \bgroup et al\mbox.\egroup
  }{2017}]{isola2017image}
{Isola}, P.; {Zhu}, J.; {Zhou}, T.; et~al.
\newblock 2017.
\newblock Image-to-image translation with conditional adversarial networks.
\newblock In {\em IEEE Conference on Computer Vision and Pattern Recognition},
  5967--5976.

\bibitem[\protect\citeauthoryear{{Jégou} \bgroup et al\mbox.\egroup
  }{2017}]{jegou2017one}
{Jégou}, S.; {Drozdzal}, M.; {Vazquez}, D.; et~al.
\newblock 2017.
\newblock The one hundred layers tiramisu: Fully convolutional densenets for
  semantic segmentation.
\newblock In {\em IEEE Conference on Computer Vision and Pattern Recognition
  Workshops},  1175--1183.

\bibitem[\protect\citeauthoryear{{Long}, {Shelhamer}, and
  {Darrell}}{2015}]{Long2015}
{Long}, J.; {Shelhamer}, E.; and {Darrell}, T.
\newblock 2015.
\newblock Fully convolutional networks for semantic segmentation.
\newblock In {\em IEEE Conference on Computer Vision and Pattern Recognition},
  3431--3440.

\bibitem[\protect\citeauthoryear{Mejjati \bgroup et al\mbox.\egroup
  }{2018}]{mejjati2018unsupervised}
Mejjati, Y.~A.; Richardt, C.; Tompkin, J.; et~al.
\newblock 2018.
\newblock Unsupervised attention-guided image-to-image translation.
\newblock In {\em Advances in Neural Information Processing Systems},
  3693--3703.

\bibitem[\protect\citeauthoryear{{Menze} \bgroup et al\mbox.\egroup
  }{2015}]{Menze2015Brain}
{Menze}, B.~H.; {Jakab}, A.; {Bauer}, S.; et~al.
\newblock 2015.
\newblock The multimodal brain tumor image segmentation benchmark.
\newblock {\em IEEE Transactions on Medical Imaging} 34(10):1993--2024.

\bibitem[\protect\citeauthoryear{{Nie} \bgroup et al\mbox.\egroup
  }{2018}]{Nie2018}
{Nie}, D.; {Trullo}, R.; {Lian}, J.; et~al.
\newblock 2018.
\newblock Medical image synthesis with deep convolutional adversarial networks.
\newblock {\em IEEE Transactions on Biomedical Engineering} 65(12):2720--2730.

\bibitem[\protect\citeauthoryear{Ronneberger, Fischer, and
  Brox}{2015}]{ronneberger2015u}
Ronneberger, O.; Fischer, P.; and Brox, T.
\newblock 2015.
\newblock U-net: Convolutional networks for biomedical image segmentation.
\newblock In {\em Medical Image Computing and Computer-Assisted Intervention},
  234--241.

\bibitem[\protect\citeauthoryear{Sharma and Hamarneh}{2019}]{sharma2019missing}
Sharma, A., and Hamarneh, G.
\newblock 2019.
\newblock Missing {MRI} pulse sequence synthesis using multi-modal generative
  adversarial network.
\newblock {\em arXiv preprint arXiv:1904.12200}.

\bibitem[\protect\citeauthoryear{{Soleymanifard} and
  {Hamghalam}}{2019}]{Soleymanifard}
{Soleymanifard}, M., and {Hamghalam}, M.
\newblock 2019.
\newblock Segmentation of whole tumor using localized active contour and
  trained neural network in boundaries.
\newblock In {\em 2019 5th Conference on Knowledge Based Engineering and
  Innovation (KBEI)},  739--744.

\bibitem[\protect\citeauthoryear{Wolterink \bgroup et al\mbox.\egroup
  }{2017}]{wolterink2017deep}
Wolterink, J.~M.; Dinkla, A.~M.; Savenije, M.~H.; et~al.
\newblock 2017.
\newblock Deep {MR} to {CT} synthesis using unpaired data.
\newblock In {\em Simulation and Synthesis in Medical Imaging},  14--23.

\bibitem[\protect\citeauthoryear{{Zhang}, {Yang}, and
  {Zheng}}{2018}]{Zhang2018}
{Zhang}, Z.; {Yang}, L.; and {Zheng}, Y.
\newblock 2018.
\newblock Translating and segmenting multimodal medical volumes with cycle- and
  shape-consistency generative adversarial network.
\newblock In {\em IEEE/CVF Conference on Computer Vision and Pattern
  Recognition},  9242--9251.

\bibitem[\protect\citeauthoryear{Zhao \bgroup et al\mbox.\egroup
  }{2017}]{zhao2017supervoxel}
Zhao, C.; Carass, A.; Lee, J.; et~al.
\newblock 2017.
\newblock A supervoxel based random forest synthesis framework for
  bidirectional mr/ct synthesis.
\newblock In {\em International Workshop on Simulation and Synthesis in Medical
  Imaging},  33--40.

\bibitem[\protect\citeauthoryear{{Zhu} \bgroup et al\mbox.\egroup
  }{2017}]{zhu2017unpaired}
{Zhu}, J.; {Park}, T.; {Isola}, P.; et~al.
\newblock 2017.
\newblock Unpaired image-to-image translation using cycle-consistent
  adversarial networks.
\newblock In {\em IEEE International Conference on Computer Vision},
  2242--2251.

\end{thebibliography}
\bibliographystyle{aaai}
\end{document}